\documentclass[11pt,a4paper]{article}

\usepackage{amsmath}
\usepackage{amssymb}
\usepackage{units}
\usepackage{slashed}

\usepackage{a4wide}
\usepackage{cite}

\usepackage{graphicx}
\usepackage{psfrag}
\usepackage{subfig}
\usepackage{feynmp}

\usepackage{booktabs}
\usepackage{multirow}
\usepackage{rotating}
\usepackage{dcolumn}

\newcommand{\into}{\ensuremath{\rightarrow}}%{\ensuremath{\;\rightarrow\;}}
\newcommand{\software}[1]{\texttt{\uppercase{#1}}}
\newcommand{\PTj}{HT'}%{P_T(j_1,j_2)}%{P_T(2j)}%{P_T^{2j}}

\title{
\vspace{-4.5ex}
{\normalsize \raggedright
DESY 11--221\\
November 2011\\[10ex]
}
\textbf{Searching for light higgsinos\\ with $\mathbf b$-jets and missing leptons}
\vspace{2ex}
}

\author{S.~Bobrovskyi, F.~Br\"ummer, W.~Buchm\"uller and J.~Hajer\\[1ex]
\textit{\normalsize Deutsches Elektronen-Synchrotron DESY,}\\
\textit{\normalsize Notkestra\ss e 85, D-22607 Hamburg, Germany}
\vspace{3ex}
}

\date{}

\begin{document}

\begin{fmffile}{feyn}

\maketitle

\thispagestyle{empty}

\begin{abstract}
\noindent A recently proposed class of supersymmetric models predicts rather light and nearly mass-degenerate higgsinos, while the other superparticles are significantly heavier. In this paper we study the early LHC phenomenology of a benchmark model of this kind. If the squarks and gluinos, and in particular the lighter stop, are still light enough to be within reach, then evidence for our model can be found in hadronic SUSY searches. Moreover, with dedicated searches it will be possible to distinguish the light higgsino model from generic SUSY models with a bino LSP. Search channels with $b$-jets and with isolated leptons play a crucial role for model discrimination.

\end{abstract}

\clearpage

\section{Introduction}

Recently a class of supersymmetric models was proposed \cite{Brummer:2011yd}
whose most characteristic feature is a large separation between the higgsino
masses and the masses of the other superparticles. The particle content is
that of the MSSM. Two higgsino-like neutralinos and a higgsino-like chargino are
light: Their masses can be arbitrarily close to the direct chargino search bound
from LEP, $m_{\chi^\pm_1}\gtrsim \unit[105]{GeV}$. There is also a light
Standard Model-like Higgs around \unit[120]{GeV}. The heavier Higgs bosons, as well as
the gaugino-like neutralinos and chargino, gluino, squarks and sleptons, have
masses of at least \unit[500]{GeV}, and possibly ranging up to several \unit{TeV} (depending
on the model details).

This kind of spectrum is naturally compatible with both LHC and LEP search bounds.
Concerning the LHC, the absence of any signals for supersymmetry in cascade
decays of first-generation squarks and gluinos points to them being rather heavy.
On the other hand, evading the LEP bound on the lightest Higgs mass requires large
loop corrections from third-generation soft terms, at least within the MSSM.
This points to large third-generation squark masses (bringing with them the
inevitable fine-tuning which is present in the remaining parameter regions of the
MSSM). Charginos and neutralinos, by contrast, can comfortably have masses between
around only \unit[100 -- 200]{GeV}.

In the present paper we investigate the discovery potential of this scenario at the early LHC.
If the only kinematically accessible states are indeed the light higgsinos
(which is the case that was recently studied in some detail in \cite{Baer:2011ec}),
the prospects tend to be poor.
While higgsinos will be abundantly produced in electroweak processes at the LHC,
any signals from their decays will be overwhelmed by Standard Model backgrounds.
We therefore focus on the less extreme case where the mass separation between
higgsinos and coloured superparticles is large,
but not so large that the latter are altogether out of reach.

The lightest coloured superparticle in the models of \cite{Brummer:2011yd} is
always the right-handed scalar top quark. Stops decaying into higgsinos will
consequently play the most important role in our study.\footnote{For some recent
related studies of stops at the LHC, see for example \cite{Bornhauser:2010mw,
Papucci:2011wy, Bi:2011ha, Desai:2011th, Brust:2011tb}.} Because of the large mass
difference, this decay will give rise to very high-energetic jets, which provides
a handle to distinguish signal events from Standard Model backgrounds. On the other
hand, we can also discriminate between our light higgsino scenario and a generic
MSSM with comparable squark and gluino masses. This is because in the latter one
would expect to see also events with high-energetic isolated leptons from
chargino and neutralino decays. Such events are absent in our scenario, since the
higgsino-like chargino and neutralinos are nearly degenerate in mass; consequently,
leptons in the final state are too soft to be detected.

This work is organized as follows. In Section~\ref{sec:model} we briefly review the
class of models proposed in \cite{Brummer:2011yd}, and present the benchmark
spectrum which our analysis will be based on. Section \ref{sec:signatures} contains
a survey of the expected collider signatures. We comment on our analysis method
in Section \ref{sec:method}, and present the results of our study in Section
\ref{sec:results}. Finally, Section 6 contains our conclusions.

\section{Light higgsinos from higher-dimensional GUTs}
\label{sec:model}

The MSSM with light higgsinos and otherwise heavy superparticles has
previously been studied e.g.~in \cite{Kane:1998ib}.
Recently some models were constructed which predict precisely this pattern,
such as the ``lopsided gauge mediation'' models of
\cite{Csaki:2008sr,DeSimone:2011va}, as well as the mixed gauge-gravity
mediation models of \cite{Brummer:2011yd} which our analysis will be concerned
with.

To briefly motivate the models of \cite{Brummer:2011yd}, we start by observing
that, in certain grand-unified models which naturally emerge from string
constructions, there is a large number of vector-like states in incomplete GUT
multiplets which should decouple close to the GUT scale. They serve as
messengers for gauge-mediated supersymmetry breaking, inducing gaugino masses
and scalar soft masses. Because of the high mediation scale, gravity mediation
cannot be neglected. The gravity-mediated contributions to the MSSM parameters
are however subdominant with respect to the gauge-mediated ones. The only
exception are the $\mu$ and $B_\mu$ parameters, to which (minimal) gauge
mediation does not contribute at all. These two parameters are induced by
gravitationally suppressed interactions through the Giudice-Masiero mechanism,
leading to light higgsinos and otherwise heavy superparticles. Related models
with mixed gauge-gravity mediation have previously been discussed in
\cite{Lalak:2008bc}.

The main properties of the higgsino sector can be summarized as follows. Since
\begin{equation}
    |\mu|
\ll |M_1|, \
    |M_2|
\ ,
\end{equation}
where $M_1$ and $M_2$ are the bino and wino masses respectively, there are three
higgsino-like light states $\chi_1^0$, $\chi_1^\pm$ and $\chi_2^0$ with masses
close to $|\mu|$. Their mass splittings will be of the order $|\mu|^2/|M_{1,2}|$,
typically a few \unit{GeV} for $|\mu| \gtrsim \unit[100]{GeV}$ and for \unit{TeV}-scale gaugino masses.
A $\chi_1^0$ LSP is not a viable dark matter candidate, since its relic density
is extremely low due to efficient chargino coannihilation. This same mechanism,
on the other hand, can substantially ameliorate the gravitino BBN problem if
dark matter consists of gravitinos instead. The $\chi^0_1$ is then the NLSP, but
it will be effectively stable on collider timescales.

The precise details of the spectrum depend on the messenger content of the
model, on the exact choice of messenger scale and SUSY breaking scale, and on
the assumptions about the gravity-mediated contributions to the soft terms. For
our purposes of a first tentative study of collider phenomenology, it is
convenient to adopt a simplified parametrization: We fix the gravitino mass to
be $m_{\nicefrac{3}{2}} = \unit[100]{GeV}$, and choose a common messenger mass just below the GUT
scale, $M_{\rm m} = \unit[5 \times 10^{15}]{GeV}$. Then the essential free parameters are
the gaugino masses $M_1$, $M_2$ and $M_3$, the Higgs soft mass mixing $B_\mu$,
and the higgsino mass $\mu$. At the GUT scale we expect
$|B_\mu| \simeq |\mu|^2 \simeq m_{\nicefrac{3}{2}}^2$ and $|M_{1,2,3}| \gg m_{\nicefrac{3}{2}}$.
Scalar soft masses are dominated by the gauge-mediated contribution, which is completely
fixed after prescribing the gaugino masses. Explicitly, they are given by the
standard minimal gauge mediation formula
\begin{equation}
    m_\Phi^2
  = 2\left(\frac{g^2}{16 \pi^2}\right)^2
    \left( \sum_a C_a \, n_a \right)
    \left| \frac{F}{M_{\mathrm m}} \right|^2
\ ,
\end{equation}
where $a=1,2,3$ labels the Standard Model gauge factors, $C_a$ is the
corresponding quadratic Casimir of $\Phi$, the SUSY breaking scale $F$ is
\begin{equation}
    F
  = \sqrt 3 \, m_{\nicefrac{3}{2}} \, M_\mathrm{Planck}
  = ( \unit[2 \times 10^{10}]{GeV})^2
\ ,
\end{equation}
and the effective messenger numbers $n_a$ are obtained by inverting the standard
gaugino mass formula
\begin{equation}
    M_a
  = \frac{g^2}{16\pi^2} n_a \, \frac{F}{M_{\mathrm m}}
\,.
\end{equation}
We are neglecting the running of the gauge couplings between $M_{\rm m}$ and $M_{\rm GUT}$, as well as the subdominant gravity-mediated contributions. Trilinear terms are again dominated by gravity mediation;
for simplicity we choose them to be universal and set $A_0 = \mu$.

Having thus fixed the MSSM parameters at the messenger scale, we evolve them to
the weak scale by means of their renormalization group equations using
\software{SOFTSUSY} \cite{Allanach:2001kg}. Reproducing the correct value of the
$Z$ mass further reduces the number of free parameters by one. In the end,
within our simplified ansatz the mass spectrum is entirely determined by the
five parameters $M_1$, $M_2$, $M_3$, $\mu$ and $B_\mu$ at the messenger scale.
These are subject to the conditions that electroweak symmetry should be broken
with $m_Z=\unit[91]{GeV}$, and that there should be a separation of mass scales
according to
\begin{equation}
    \mu
 \sim
    \sqrt{B_\mu}
  \sim
    m_{\nicefrac{3}{2}}
\; \ll \;
    M_{1}
  \sim
    M_2
  \sim
 M_3
\ .
\end{equation}

\begin{table}
\centering
\begin{tabular}{ l l r r r r r }
\toprule
 \multicolumn{2}{c}{particle} & \multicolumn{5}{c}{model} \\
 \cmidrule{3-7}
 & &  \multicolumn{1}{r}{Spectrum~I} & \multicolumn{1}{r}{Spectrum~II} & \multicolumn{1}{r}{HH50} & \multicolumn{1}{r}{HH$50'$} & \multicolumn{1}{r}{simplified} \\
\midrule
 \hspace*{3em} & $h_0$ & 116 & 121 & 115 & 117 & 117 \\
\midrule
 & $\chi^0_1$ &  124 & 117 & 206 & 207 & 125 \\
 & $\chi^\pm_1$ & 129 & 119 & 389 & 395 \\
 & $\chi^0_2$ & 134 & 121 & 389 & 395 \\
\midrule
 & $\chi^0_3$ &  559 & 1\,319 & 635 & 771 \\
 & $\chi^0_4$ & 1\,059 & 2\,453 & 649 & 778 \\
 & $\chi^\pm_2$ & 1\,059 & 2\,453 & 648 & 779 \\
\midrule
 & $H_0$ & 641 & 660 & 861 & 958 \\
 & $A_0$ & 642 & 666 & 861 & 958 \\
 & $H^\pm$ & 648 & 672 & 865 & 962 \\
\midrule
 & $\tilde g$ & 1\,063 & 2\,485 & 1\,167 & 1\,167 \\
\midrule
 & $\tilde t_1$ & 665 & 1\,558 & 860 & 660 & 659 \\
 & $\tilde b_1$ & 797 & 1\,614 & 1\,034 & 943 \\
 & $\tilde u_1$ & 1\,155 & 2\,438 & 1\,122 & 1\,130 \\
 & $\tilde d_1$ & 1\,065 & 2\,294 & 1\,119 & 1\,127 \\
 \multicolumn{2}{r}{other squarks} & 1\,070 -- 1\,500 & 2\,300 -- 3\,100 & 1\,120 -- 1\,160 & 990 -- 1\,270  \\
\midrule
 & $\tilde\tau_1$ & 509 & 669  & 528 & 520 \\
 \multicolumn{2}{r}{other sleptons} & 790 -- 1\,160 & 1\,400 -- 2\,300 & 530 -- 600 & 530 -- 600 \\
\bottomrule
\end{tabular}
\caption{A light and a heavy spectrum, with a CMSSM point HH50, a CMSSM-like point HH$50'$ and a simplified model for comparison.
The parameters defining these models are listed in Table \ref{tab:model parameter}. Particle masses are in \unit{GeV}.
}
\label{tab:spectrum}
\end{table}

Table~\ref{tab:spectrum} shows two examples for low-energy spectra, both with $\mu = \unit[150]{GeV}$ and $\sqrt{B_\mu}=\unit[200]{GeV}$ and with equal values for $M_1$ and $M_2$.
Spectrum~I has $M_1 = M_2 = \unit[1250]{GeV}$ and $M_3 = \unit[428]{GeV}$;
these parameters are chosen such that the model is close to the present LHC exclusion limits.
Spectrum~II has $M_1 = M_2 = \unit[3]{TeV}$ and $M_3 = \unit[1130]{GeV}$, for which the model would be invisible at the early LHC and quite difficult to find even at \unit[14]{TeV}.
Our analysis will be mostly concerned with the phenomenology of Spectrum~I at $\sqrt{s}=\unit[7]{TeV}$.

For comparison, we have also included a similar CMSSM benchmark point HH50 and a CMSSM-like benchmark point HH$50'$.
HH50 has $m_0=M_{1/2}=\unit[500]{GeV}$, $\tan\beta=10$, $\operatorname{sign}(\mu)=+1$ and $A_0=0$.
HH$50'$ is defined in the same way, but with the soft terms of the third generation chosen differently:
Third-generation squarks and sleptons were given a universal soft mass $m_0^{(3)}=\unit[300]{GeV}$ and a trilinear $A$-parameter $A_0^{(3)}=\unit[-1]{TeV}$.
This choice was made in order to have a reference spectrum whose
$\tilde t_1\tilde t_1^*$ production cross section is comparable to that of
Spectrum~I, while closely resembling the CMSSM. Finally, we also list a comparable
simplified model, containing only the $\tilde t_1$ and a bino-like neutralino LSP.
The model definitions are summarized in Table \ref{tab:model parameter}.

\begin{table}
\centering
\begin{tabular}{ l r r r r r r r r r r}
\toprule
 model & $\mu$ & $\sqrt{B_\mu}$ & $M_1 = M_2$ & $M_3$ & $m_0$ & $m_0^{(3)}$ & $A_0$ & $A_0^{(3)}$ & $\tan \beta$ \\
\midrule
 Spectrum~I & 150 & 200 & 1\,250\hspace*{1em} & 428 & & & & & 46 \\
 Spectrum~II & 150 & 200 & 3\,000\hspace*{1em} & 1\,130 & & & & & 53 \\ \midrule
 HH50  & & & 500\hspace*{1em} & 500 & 500 & 500 & 0 & 0 & 10 \\
 HH$50'$ & & & 500\hspace*{1em} & 500 & 500 & 300 & 0 & $-1\,000$ & 10 \\
\bottomrule
\end{tabular}
\caption{Defining parameters for a light and a heavy spectrum, with a CMSSM point and a CMSSM-like point for comparison.
Particle masses are in \unit{GeV}.
In HH$50'$ third-generation squarks and sleptons were given a universal soft mass $m_0^{(3)}$ and a trilinear $A$-parameter $A_0^{(3)}$.
}
\label{tab:model parameter}
\end{table}

\section{Signatures}
\label{sec:signatures}

\begin{figure}
\centering
\scriptsize
\psfrag{pt}{\footnotesize\hspace{-3em}$p_T$ [\unit{GeV}]}
\psfrag{Leptons}{\footnotesize\hspace{-1em}Leptons}
\psfrag{Events}{\footnotesize\hspace{-1em}Events}
\psfrag{etmis}{\footnotesize\hspace{-2.3em}$\slashed E_T$ [\unit{GeV}]}
\psfrag{WZ}{\footnotesize$WZ$}
\psfrag{Spectrum I}{\footnotesize Spectrum I}
\psfrag{0}{0}
\psfrag{1}{1}
\psfrag{2}{\tiny\hspace{.2em}2}
\psfrag{3}{\tiny\hspace{.2em}3}
\psfrag{4}{\tiny\hspace{.2em}4}
\psfrag{10}{10}
\psfrag{15}{15}
\psfrag{20}{20}
\psfrag{25}{25}
\psfrag{30}{30}
\psfrag{35}{35}
\psfrag{40}{40}
\psfrag{45}{45}
\psfrag{50}{50}
\psfrag{100}{100}
\psfrag{150}{150}
\psfrag{200}{200}
\psfrag{250}{250}
\subfloat{\includegraphics[width=0.46\textwidth]{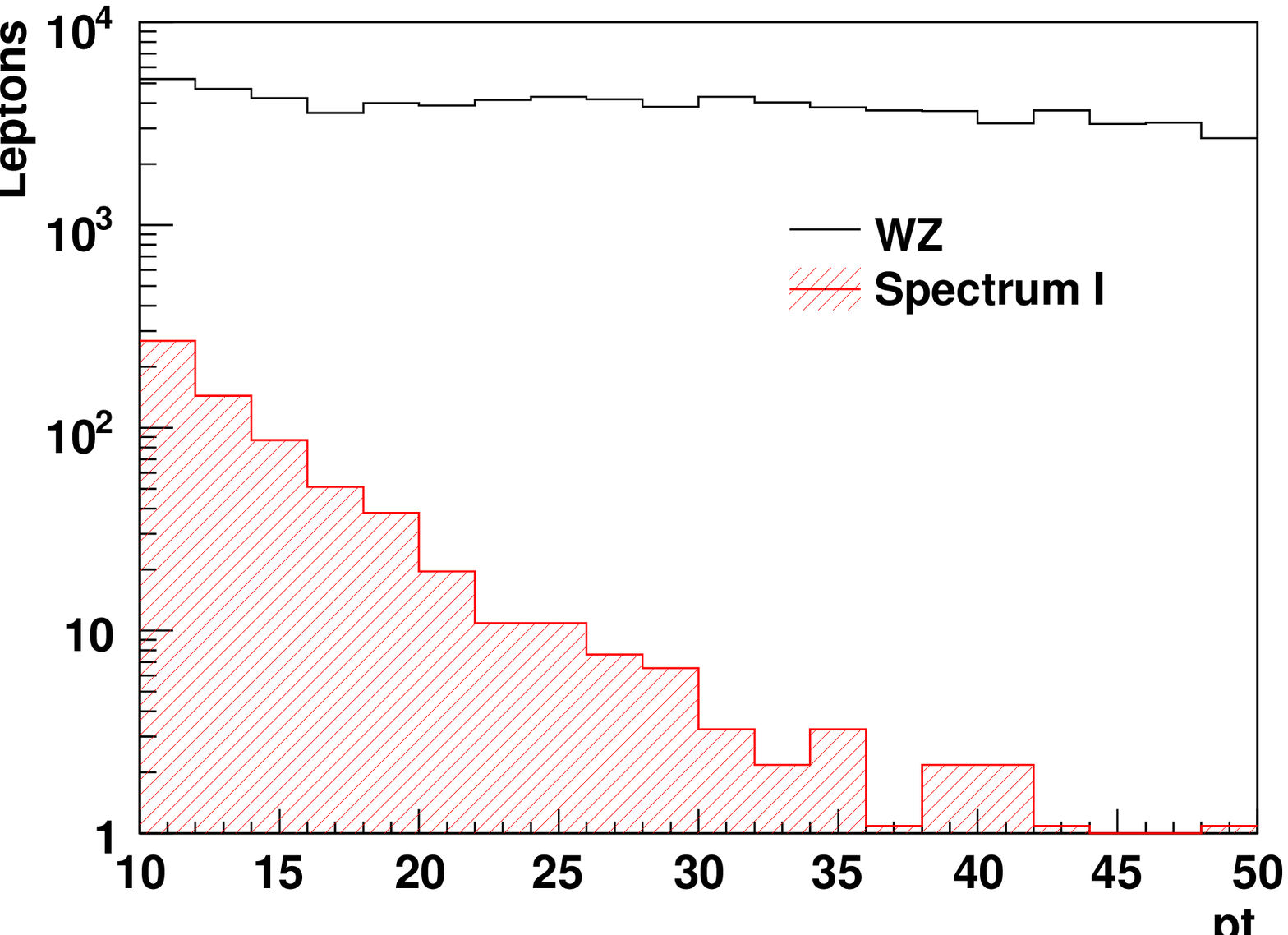}}
\hspace*{0.02\textwidth}
\subfloat{\includegraphics[width=0.49\textwidth]{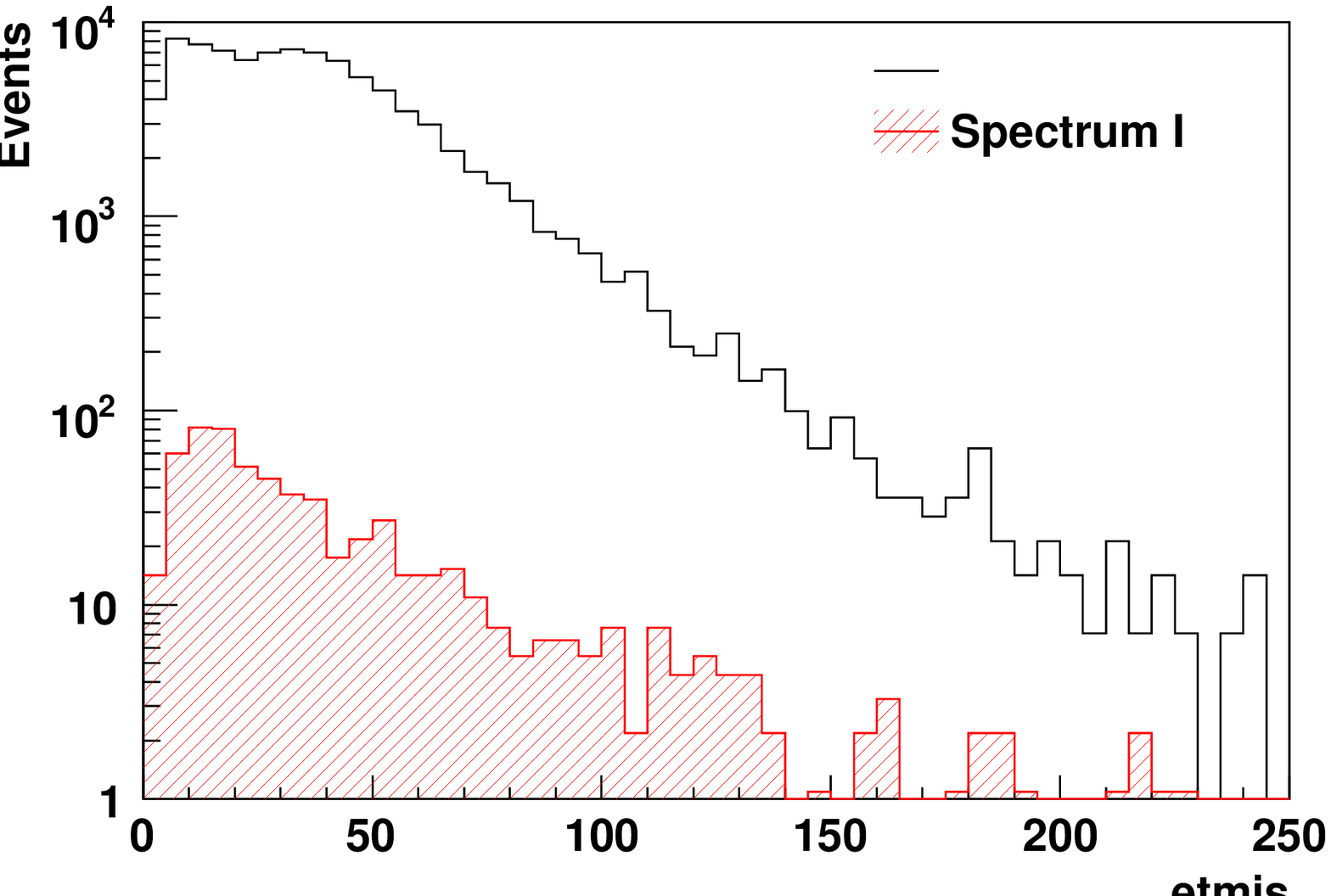}}
\caption{Lepton transverse momentum and missing transverse energy distributions of leptonic events from higgsino decays with Spectrum~I. The higgsinos are produced in electroweak processes;
the numbers are scaled to $\unit[20]{fb^{-1}}$. For comparison, the SM background from $WZ$ production (which is just one of the several contributing processes) is also shown. See Section \ref{sec:method} for details of the event simulation.}
\label{fig:lepton pT}
\end{figure}

The light higgsinos of our scenario will be produced in copious numbers in electroweak processes at the LHC.
The Drell-Yan process gives rise to $\chi_1^+ \chi_1^-$, $\chi_1^\pm \chi_{1,2}^0$ and $\chi^0_1 \chi^0_2$ final states,
and $W$ boson fusion can give like-sign $\chi^\pm_1 \chi^\pm_1$ pairs.
The subsequent decays of $\chi_2^0$ and $\chi_1^\pm$ into $\chi_1^0$ will lead to events with missing energy and soft jets or leptons.

Unfortunately, with the higgsino mass splittings in the range of only a few \unit{GeV},
most of these jets and leptons are too soft to even trigger on, and those events with high
enough $p_T$ to be detected are completely swamped by the Standard Model background. Demanding
large missing transverse energy does not help much, since also the $\slashed{E}_T$ spectrum falls very
rapidly. For illustration, the lepton $p_T$ and $\slashed{E}_T$ distributions for Spectrum~I are shown in Figure~\ref{fig:lepton pT}. We have also studied events with additional jets from initial-state gluon radiation, in order to increase the number of events with larger $p_T$ and $\slashed{E}_T$. While this somewhat enhances the tails of the distributions, it also reduces the overall cross section, and the combined effect does very little to improve the overall situation.
In conclusion we confirm the findings of \cite{Baer:2011ec} that,
in order to find evidence for our scenario in electroweak processes, a linear collider would be far better suited.\footnote{For the LHC, a monojet (from initial-state gluon radiation) together with large missing $E_T$ might perhaps be a useful signal, in combination with other searches. We will however not pursue this possibility in the present work because of the difficulties in accurately estimating the background without a full detector simulation. 
}

We are therefore led to consider those regions of parameter space where some coloured superparticles are still light enough to be produced at the LHC.
The lightest coloured superparticle in our class of models is always the lighter of the scalar top quarks $\tilde t_1$.
At the LHC it may be produced in pairs,
or it may appear in cascade decays of first-generation squarks and gluinos if these are kinematically accessible.
It turns out that processes involving the $\tilde t_1$ are particularly well suited to find evidence for our scenario (or to constrain it),
and also to distinguish it from more generic incarnations of the MSSM.

For definiteness we will from now on focus on the Spectrum~I benchmark point
$M_1 = M_2 = \unit[1250]{GeV}$, $M_3 = \unit[428]{GeV}$, $\mu = \unit[150]{GeV}$, $\sqrt{B_\mu} = \unit[200]{GeV}$.
In a sense this is a maximally optimistic set of parameters,
chosen such that it is still marginally allowed by current search limits.
We are planning to extend our analysis to cover a wider range of parameter space in the future.

With superparticle masses as in Spectrum~I, the clearest signatures at the early LHC will be jets with missing $E_T$.
We will see that the cross sections for stop pair production on the one hand and the more familiar $\tilde q\tilde q$, $\tilde q\tilde q^*$, $\tilde q\tilde g$ and $\tilde g\tilde g$ production (where $\tilde q$ stands for any first-generation squark) on the other hand are comparable; all these processes contribute to the signal.

\begin{figure}
\centering
\parbox[c][115pt][c]{120pt}{
\begin{fmfgraph*}(120,100)
 \fmfpen{.5pt}
 \fmfset{arrow_len}{2mm}
 \fmfset{arrow_len}{2mm}
 \fmfset{curly_len}{2mm}
 \fmfleft{i1}
 \fmfright{o1,o2,o3,o4}
 \fmf{gluon,tension=2}{i1,v1}
 \fmf{dashes,label=$\tilde t$,label.side=right}{v1,v2}
 \fmf{dashes,label=$\tilde t^*$,label.side=left}{v1,v3}
 \fmf{fermion}{v2,o1}
 \fmf{fermion}{o2,v2}
 \fmf{fermion}{v3,o3}
 \fmf{fermion}{o4,v3}
 \fmflabel{$\chi^-_1$}{o3}
 \fmflabel{$\chi^0_{1,2}$}{o2}
 \fmflabel{$\bar b$}{o4}
 \fmflabel{$t$}{o1}
 \fmflabel{$g$}{i1}
\end{fmfgraph*}
}
\mbox{$\quad\; \left.\begin{array}{c} \\ \\ \\ \\ \end{array}\right\}$ (invisible)}
\caption{An example for a stop pair production event, showing up as to two high-energetic $b$-jets and missing energy.}
\label{fig:stoppair}
\end{figure}
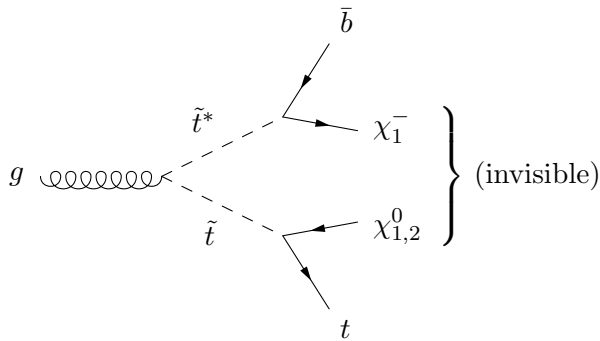

More importantly, once there is evidence for supersymmetry in searches for jets plus missing $E_T$,
our model can also be distinguished experimentally from generic variants of the MSSM which lack its characteristic features of light and near-degenerate higgsinos.
This is achieved by focussing on the stop pair production channel.
In Spectrum~I, stop decays do not involve hard leptons,
since possible leptons from $\chi^0_2$ or $\chi^\pm_1$ decays are too soft to be detected.
The signature of a $\tilde t_1$ is therefore always a hard $b$-jet plus missing $E_T$;
a typical stop pair event is shown in Figure~\ref{fig:stoppair}.
By contrast, in generic supersymmetric models one usually expects also events with jets, missing $E_T$ and isolated leptons,
be it from cascade decays of squarks and gluinos or from $\tilde t$ decaying into charginos or non-LSP neutralinos.
Once a signal is found in the jets + MET channel,
we could use the absence of signals with leptons to severely constrain interpretations in terms of generic supersymmetry,
thus providing further indirect evidence for our scenario.

We may even be able to discriminate between our model and a ``simplified model'' comprising only a $\tilde t_1$ and a bino-like $\chi^0_1$.
In such a framework, likewise, no events with hard isolated leptons are expected.
However, because the only possible $\tilde t_1$ decay is then $\tilde t_1\into t\chi^0_1$ with the $t$ decaying further into $bW$,
the $b$-jet spectrum turns out to be significantly different from that of our model,
where about half of the stops decay directly into a $b$ quark without an intermediate top.

In the following Sections we present the results of three simulated searches.
The first is for jets and large missing $E_T$, in order to show that early LHC will be able to find evidence for our model. The second also includes leptons, to show that early LHC will, furthermore, be able to distinguish our model from a comparable CMSSM-like model.
More precisely, our model will be compared both with the CMSSM point HH50, which has similar $\tilde g$ and $\tilde q$ production cross sections,
and with the CMSSM-like point HH$50'$, which in addition has also a comparable $\tilde t_1$ pair production cross section.
Finally, we present a search with the cuts optimized to select events from $\tilde t_1$ pair production, and compare the result with the simplified model mentioned above.

\section{Simulation of signal and background}
\label{sec:method}

\begin{table}
\centering
\begin{tabular}{ l D{.}{.}{-1} D{.}{.}{-1} D{.}{.}{-1} D{.}{.}{-1} D{.}{.}{-1} D{.}{.}{-1} D{.}{.}{-1} }
\toprule
   model & \multicolumn{1}{r}{$\sigma (\tilde{q} \tilde{q}^*)$} & \multicolumn{1}{r}{$ \sigma (\tilde{q}\tilde{q})$}
 & \multicolumn{1}{r}{$\sigma (\tilde{t} \tilde{t}^*)$} & \multicolumn{1}{r}{$\sigma (\tilde{b} \tilde{b}^*)$}
 & \multicolumn{1}{r}{$\sigma (\tilde{g} \tilde{g})$} & \multicolumn{1}{r}{$\sigma (\tilde{g} \tilde{q})$} & \multicolumn{1}{r}{$\sigma (\text{tot})$} \\
\midrule
 Spectrum~I & 0.388 & 3.83 & 5.61 & 0.6 & 2.9 & 8.45 & 21.78 \\
 HH 50 & 1.79 & 12 & 0.682 & 0.044 & 1 & 9.3 & 24.8 \\
 HH 50' & 1.65 & 11.5 & 5.96 & 0.136 & 0.979 & 8.9 & 29 \\
\bottomrule
\end{tabular}
\caption{Production cross sections of different models in \unit{fb} calculated with \software{PROSPINO}.
The cross section for  $\tilde{b} \tilde{b}^*$-production is given at the lowest order,
all other cross sections are calculated at NLO.}
\label{tab:sigcross}
\end{table}

All Monte Carlo samples were generated with \software{MADGRAPH 4.4.44} \cite{Alwall:2007st} interfaced with \software{PYTHIA 6.4.22} \cite{Sjostrand:2006za} using \software{CTEQ6L1} parton distribution functions  \cite{Pumplin:2002vw}.
In order to generate signal events, decay widths of supersymmetric particles were computed with \software{SDECAY} \cite{Muhlleitner:2004mka} from spectra calculated with \software{SOFTSUSY}.
The generic detector simulation \software{DELPHES} \cite{Ovyn:2009tx}, tuned to the CMS detector, was used in order to account for effects of event reconstruction at the detector level.

The signal production cross sections are listed in Table~\ref{tab:sigcross}.
For Spectrum~I $43\,500$ signal events were simulated, to be compared with $435$ events expected at the early LHC with an assumed integrated luminosity yield of $\unit[20]{fb^{-1}}$.
The number of signal events passing the cuts should therefore eventually be divided by a normalization factor $100$ for a realistic estimate. For HH50 and HH$50'$, we simulated $10\,000$ events each, with respectively $496$ and $580$ events expected, so the normalization factors are $20$ and $17$ respectively.

The corresponding figures for Standard Model backgrounds are listed in Table~\ref{tab:backcross}.
It turns out that $t\bar t$ is the most important background. Since, consequently,
the best statistics is needed for this channel, we have simulated about three times more events than expected. For the remaining backgrounds, the number of simulated events roughly matches the number of expected events, or exceeds it in the case of tri-bosons (where the cross sections are small) in order to avoid large Monte Carlo errors. An exception are background events with vector bosons plus jets, where we have only simulated a small fraction of the expected events. However, as will become clear when we present the cut flows, this background is very efficiently removed by our cuts. Therefore it can be safely neglected without having to simulate the full sample.

\begin{table}
\centering
\begin{tabular}{ l r r r}
\toprule
 sample & \multicolumn{1}{c}{$\sigma$ in \unit{pb}} & \multicolumn{2}{c}{events} \\
\cmidrule{3-4}
 & & \multicolumn{1}{c}{expected}  & \multicolumn{1}{c}{simulated}\\
\midrule
 $t \bar{t}$ & 163 & $3.3 \times 10^6$ & $11.3 \times 10^6$\\
 single top & 85.1 & $1.7 \times 10^6$ & $1.7 \times 10^6$ \\
 $W+\text{jet}$ & 826 & $1\,652 \times 10^4$ & $5 \times 10^4$  \\
 $W^+W^-$ & 44.974 & $899 \times 10^3$ & $1\,000 \times 10^3$  \\
 $W^+Z$ & 11.580 & \multirow{2}{*}{$\Bigl\}\quad 358 \times 10^3$} & \multirow{2}{*}{$400 \times 10^3$} \\
 $W^-Z$ & 6.342 \\
 $ZZ$ & 6.195 & $124 \times 10^3$ & $150 \times 10^3$ \\
 $W^+W^-W^+$ & $4 \times 10^{-2}$ & 800 & $15\,000$ \\
 $W^+W^-Z$ & $3 \times 10^{-2}$ & 600 & $15\,000$ \\
 $W^+ZZ$ & $9 \times 10^{-3}$ & 180 & $15\,000$ \\
 $ZZZ$ & $3 \times 10^{-3}$ & 60 & $5\,629$\\
\bottomrule
\end{tabular}
\caption{Cross sections and numbers of generated events of SM background used in the present analysis.
The single top production cross-section includes all LHC production channels.
The cross sections for the tri-boson events are calculated at the Born level with \software{MADGRAPH},
all other cross sections are taken from  \cite{Kidonakis:2010bb,Campbell:2011bn,Torrielli:2010aw}}
\label{tab:backcross}
\end{table}

\section{Event selection and analysis}
\label{sec:results}

\subsection{Discovery with all-hadronic search}\label{sec:allhad}

The first analysis serves to show that LHC will be able to find evidence for our model, i.e.~to distinguish its signatures from the Standard Model background.

In the first stage, candidate events with multiple high-energetic jets and missing transverse energy are selected with the following pre-selection cuts at the level of the detector simulation:
\begin{itemize}
 \item $1<N(j)<5$ , where \quad $p_T(j) > \unit[100]{GeV}$,
 \item $\slashed{E}_T > \unit[50]{GeV}$.
\end{itemize}
Furthermore, all events with an isolated lepton (electron or muon) with $p_T > \unit[10]{GeV}$ are rejected in order to suppress events with genuine missing energy from neutrinos:
\begin{itemize}
\item $N(l) = 0$.
\end{itemize}

After imposing these pre-selection cuts, we use a set of cuts optimized for discriminating between signal and background. Events are required to satisfy
\begin{itemize}
 \item $\PTj > \unit[500]{GeV}$,
\end{itemize}
where $\PTj$ is the sum of the transverse momenta of the two most energetic jets,
\begin{equation}
\PTj=\sum_{i=1}^2 p_T(j_i)\,.
\end{equation}
Following the experimental analyses, we use the $\alpha_T$ variable \cite{Randall:2008rw, CMS-PAS-SUS-08-005, CMS-PAS-SUS-09-001} as the main discriminator against QCD multi-jet production, defined for di-jet events as:
\begin{equation}
    \alpha_T
  = \frac{E_T(j_2)}{M_T}
  = \frac{E_T(j_2)}{\sqrt{
    \left( \sum_{i=1}^2 E_T(j_i) \right)^2
  - \left( \sum_{i=1}^2 p_x(j_i) \right)^2
  - \left( \sum_{i=1}^2 p_y(j_i) \right)^2
    }}
\ ,
\end{equation}
where $j_2$ denotes the next-to-leading jet.
In our analysis we use $p_T$ of the jets provided by \software{Delphes} instead of $E_T$, and require the event to have
\begin{itemize}
 \item $\alpha_T > 0.55$
\end{itemize}
in order to pass the cut.
In events with jet multiplicity $N(j) >2$, two pseudo jets are formed following the CMS strategy \cite{CMS-PAS-SUS-09-001} and the $\alpha_T$ variable is constructed from the pseudojets.
Finally, in order to further suppress the  $t\bar{t}$ background, we demand a very high value of missing transverse energy:
\begin{itemize}
 \item $\slashed{E}_T > \unit[400]{GeV}$.
\end{itemize}
Because of the high $\slashed E_T$ cut in combination with the selection based on $\alpha_T$, we can safely neglect QCD di- and multi-jet background contributions. The resulting cut flow is shown in Table~\ref{tab:cutflow1}.

\begin{table}
\centering
\begin{tabular}{ c l r r r r r r r r }
\toprule
 & & \multicolumn{1}{l}{\hspace*{.8em}before} & \multicolumn{3}{c}{pre-cuts} & & \\
\cmidrule{4-6}
 & & \multicolumn{1}{l}{\hspace*{.8em}cuts} & \multicolumn{1}{c}{$N(j)$} & \multicolumn{1}{c}{$\slashed E_T$} & \multicolumn{1}{c}{$N(l)$} & \multicolumn{1}{c}{$\PTj$} & \multicolumn{1}{c}{$\alpha_T$} & \multicolumn{1}{c}{$\slashed E_T$} \\
\midrule
 \multirow{6}{*}{\begin{sideways}Spectrum~I\end{sideways}} & $\tilde{q} \tilde{q}^*$ & 720 & 569 & 555 & 420 & 401 & 86 & 78 \\
 & $\tilde{q} \tilde{q}$ & 7\,660 & 6\,416 & 6\,329 & 4\,788 & 4\,581 & 919 & 761 \\
 & $\tilde{t} \tilde{t}^*$ & 11\,220 & 8\,909 & 8\,729 & 7\,690 & 5\,123 & 1\,074 & 864 \\
 & $\tilde{b} \tilde{b}^*$ & 1\,200 & 993 & 983 & 866 & 691 & 162 & 138 \\
 & $\tilde{g} \tilde{g}$ & 5\,800 & 4\,678 & 4\,622 & 3\,573 & 3\,250 & 809 & 631 \\
 & $\tilde{g} \tilde{q}$ & 16\,900 & 13\,425 & 13\,257 & 10\,237 & 9\,655 & 2\,080 & 1\,685 \\
\cmidrule{4-6}\cmidrule{9-9}
 \multicolumn{2}{r}{weighted events} & & & & & & & \textbf{42} \\
\midrule
 HH50 & & 10\,000 & 8\,892 & 8\,822 & 7\,119 & 6\,882 & 1\,888 & 1\,770 \\
\cmidrule{9-9}
 \multicolumn{2}{r}{weighted events} & & & & & & & \textbf{88} \\
\midrule
 HH$50'$ & & 10\,000 & 8\,778 & 8\,691 & 6\,850 & 6\,244 & 1\,582 & 1\,467 \\
\cmidrule{9-9}
 \multicolumn{2}{r}{weighted events} & & & & & & & \textbf{84} \\
\midrule
 \multirow{5}{*}{\begin{sideways}SM\end{sideways}}& $t \bar t$ & $11.3 \times 10^6$ & $3.2 \times 10^6$ & 930\,000 & 510\,000 & 59\,992 & 312 & 64 \\
 & $t$ & $1.7 \times 10^6$ & 160\,197 & 23\,773 & 15\,089 & 2\,062 & 6 & 3 \\
 & $W+\text{jet}$ & 50\,000 & 120 & 5 & 2 & 0 & 0 & 0 \\
 & di-bosons & $1.55 \times 10^6$ & 36\,862 & 3\,820 & 2\,281 & 404 & 4 & 3 \\
 & tri-bosons & 50\,629 & 9\,051 & 2\,763 & 1\,714 & 470 & 9 & 6 \\
\cmidrule{4-6}\cmidrule{9-9}
 \multicolumn{2}{r}{weighted events} & & & & & & & \textbf{25} \\
\bottomrule
\end{tabular}
\caption{Cut flow of general all-hadronic analysis for different signals and backgrounds at $\sqrt s = \unit[7]{TeV}$.
Figures are given for all events that were simulated. The bold numbers are the events surviving all cuts, properly normalized to an integrated luminosity of $\unit[20]{fb^{-1}}$.
The cut flow for the Spectrum~I is shown separately for each different production channel.
}
\label{tab:cutflow1}
\end{table}

Evidently, with this analysis it will be possible to discriminate between our model and the Standard Model background.
The same is true for the HH50 and HH$50'$ models.
This result is of course unsurprising, since all these benchmark points were chosen to lie near the $\unit[1]{fb^{-1}}$ exclusion bounds, and here we are assuming a data sample of $\unit[20]{fb^{-1}}$.

\subsection{Model discrimination: CMSSM-like models}
\label{sec:sls}

The more interesting question is that of model discrimination.
For this a fully hadronic search such as the one we just presented is not suitable, even though the number of events passing the above cuts is significantly different between our model and HH50 / HH$50'$.
This difference could, after all, be accounted for by slightly different squark and gluino production cross sections -- for instance, the HH50 and HH$50'$ spectra would just need to be slightly heavier
in order to reproduce the 42 events after cuts which we found for our model.

In fact, some information can be gained already by requesting, in addition to the cuts of Section \ref{sec:allhad}, that at least one jet should be $b$-tagged.
We assume a $p_T$-independent $b$-tagging efficiency of \unit[40]{\%}, and a mistagging probability of \unit[10]{\%} as implemented in \software{Delphes}.
The additional cut is then
\begin{itemize}
 \item $N(b\text{-jets})\geq 1$\,.
\end{itemize}
The cut flow is shown in Table~\ref{tab:cutflow2}.
Note that the number of events from both HH50 and HH$50'$ is dramatically reduced.
This is partly because, in our model, a sizeable fraction of events was due to $\tilde t$ pair production, and the gluino can only decay into $\tilde t_1$ or $\tilde b_1$.
By contrast, in HH50 and HH$50'$ most events involve $\tilde q$ decays which do not necessarily lead to $b$-jets.
Moreover, by vetoing events with isolated leptons, fewer $\tilde t_1$ events in our model are cut away than in HH50 and HH$50'$ -- these models tend to produce more leptonic events, which we will now put to use in a separate semi-leptonic analysis.

\begin{table}
\centering
\begin{tabular}{ l l r r r r r }
\toprule
 & & \parbox{3.6em}{\raggedright After\\pre-cuts}  & \multicolumn{1}{c}{$b$-tag} & \multicolumn{1}{c}{$\PTj$} & \multicolumn{1}{c}{$\alpha_T$} & \multicolumn{1}{c}{$\slashed E_T$} \\
\midrule
 \multirow{6}{*}{\begin{sideways}Spectrum~I\end{sideways}}& $\tilde q \tilde q^*$ & 420 & 78 & 77 & 0 & 0 \\
 & $\tilde q \tilde q$ & 4\,788 & 1\,153 & 1\,126 & 226 & 183 \\
 & $\tilde t \tilde t^*$ & 7\,690 & 3\,851 & 3\,268 & 834 & 562 \\
 & $\tilde b \tilde b^*$ & 866 & 445 & 405 & 112 & 87 \\
 & $\tilde g \tilde g$ & 3\,573 & 1\,843 & 1\,793 & 465 & 351 \\
 & $\tilde g \tilde q$ & 10\,237 & 3\,940 & 3\,862 & 845 & 652 \\
\cmidrule{7-7}
 \multicolumn{2}{r}{weighted events} & & & & & \textbf{18} \\
\midrule
 \multicolumn{2}{l}{HH50} & 7119 & 631 & 619 & 124 & 108 \\
\cmidrule{7-7}
 \multicolumn{2}{r}{weighted events} & & & & & \textbf 5 \\
\midrule
 \multicolumn{2}{l}{HH$50'$} & 6\,850 & 930 & 841 & 158 & 124 \\
\cmidrule{7-7}
 \multicolumn{2}{r}{weighted events} & & & & & \textbf 7 \\
\midrule
 \multirow{2}{*}{\begin{sideways}SM\end{sideways}} & $t \bar t$ & $51 \times 10^4$ & $20 \times 10^4$  & 48\,624 & 391 & 25 \\
 & $t$  & 15\,089 & 4\,798 & 656 & 3 & 2 \\
\cmidrule{7-7}
 \multicolumn{2}{r}{weighted events} & & & & & \textbf 9 \\
\bottomrule
\end{tabular}
\caption{Cut flow of the hadronic analysis with $b$-tagging for different signals and the relevant backgrounds at $\sqrt s = \unit[7]{TeV}$.
The remaining signal and background events, scaled to an integrated luminosity of $\unit[20]{fb^{-1}}$, are printed in bold.
The cut flow for Spectrum~I is shown separately for each different production channel.
}
\label{tab:cutflow2}

\end{table}

More precisely, as explained in Section~\ref{sec:signatures}, $\tilde t_1$ decays in our model can give hard isolated leptons at most from secondary top decays
(which is, incidentally, also true for $\tilde b_1$ and even $\tilde g$ decays, since the gluino can only decay into $\tilde t_1$ or $\tilde b_1$).
In HH50 and HH$50'$ many more leptons are expected, jets with missing $E_T$ and isolated leptons being one of the hallmark signatures for generic supersymmetry. This motivates a semi-leptonic search for better model discrimination.

An event is selected for further analysis if it contains exactly one lepton (muon or electron) candidate
\begin{itemize}
 \item $N(l) = 1$ , \quad  $p_T(l) > \unit[15]{GeV}$.
\end{itemize}
Other than that, our pre-selection cuts are as before,
\begin{itemize}
 \item $N(j) > 1$ , \quad  $p_T(j) > \unit[100]{GeV}$,
 \item $\slashed{E}_T > \unit[50]{GeV}$.
\end{itemize}

The actual cuts are now as follows.
We select events with exactly two high-energetic jets,
\begin{itemize}
\item $N(j) = 2$.
\end{itemize}
This criterion selects preferably the $\tilde t\tilde t^*$ production channel, since usually more than two jets are expected to appear in channels involving $\tilde q$ or $\tilde g$. Furthermore, we employ the transverse mass variable
\begin{equation}
    m_T
  = \sqrt{ 2 p_T(l) \slashed E_T \left (1 - \cos \Delta \phi(l, \slashed E_T) \right) }
\ ,
\end{equation}
where $\Delta \phi(l, \slashed E_T)$ is the angle between missing transverse energy and the momentum of the lepton in the transverse plane.
This variable is bounded by the $W$-boson mass if the lepton and $\slashed E_T$ originate in $W$-boson decay.
We select events with
\begin{itemize}
\item $m_T> \unit[100]{GeV}$,
\end{itemize}
and ensure that the leptons in these events are isolated.
Furthermore, as in the previous analysis we demand that the two jets have high transverse momentum and high missing transverse energy,
\begin{itemize}
\item $\PTj > \unit[500]{GeV}$,
\item $\slashed{E}_T > \unit[400]{GeV}$.
\end{itemize}
\begin{table}
\centering
\begin{tabular}{ l l r r r r r r r r r }
\toprule
 & & \multicolumn{1}{l}{\hspace*{.8em}before} & \multicolumn{3}{c}{pre-cuts} & & & & \\
\cmidrule{4-6}
 & & \multicolumn{1}{l}{\hspace*{.8em}cuts} & \multicolumn{1}{c}{$N(l)$} & \multicolumn{1}{c}{$N(j)$} & \multicolumn{1}{c}{$\slashed E_T$} & \multicolumn{1}{c}{$N(j)$} & \multicolumn{1}{c}{$m_T$} & \multicolumn{1}{c}{Iso} & \multicolumn{1}{c}{$\PTj$} & \multicolumn{1}{c}{$\slashed E_T$} \\
\midrule
 \multirow{6}{*}{\begin{sideways}Spectrum~I\end{sideways}}& $\tilde{q} \tilde{q}^*$ & 720 & 238 & 233 & 229 & 26 & 17 & 6 & 6 & 1 \\
 & $\tilde{q} \tilde{q}$ & 7\,660 & 2\,690 & 2\,650 & 2\,622 & 380 & 271 & 129 & 123 & 74 \\
 & $\tilde{t} \tilde{t}^*$ & 11\,220 & 4\,063 & 3\,202 & 3\,135 & 2\,191 & 1\,701 & 230 & 90 & 40 \\
 & $\tilde{b} \tilde{b}^*$ & 1\,200 & 449 & 367 & 367 & 244 & 180 & 25 & 16 & 8 \\
 & $\tilde{g} \tilde{g}$ & 5\,800 & 2\,224 & 2\,202 & 2\,173 & 258 & 207 & 53 & 46 & 29 \\
 & $\tilde{g} \tilde{q}$ & 16\,900 & 6\,397 & 6\,346 & 6\,261 & 690 & 536 & 170 & 142 & 76 \\
\cmidrule{4-6}\cmidrule{11-11}
 \multicolumn{2}{r}{events} & & & & & & & & & \textbf 2 \\
\midrule
 \multicolumn{2}{l}{HH50} & 10\,000  & 2\,432  & 2\,352 & 2\,330 & 615 & 438 & 242 & 225 & 147 \\
\cmidrule{4-6}\cmidrule{11-11}
 \multicolumn{2}{r}{events} & & & & & & & & & \textbf 7 \\
\midrule
 \multicolumn{2}{l}{HH$50'$} & 10\,000 & 2\,699 & 2\,519 & 2\,496 & 796 & 576 & 308 & 246 & 147 \\
\cmidrule{4-6}\cmidrule{11-11}
 \multicolumn{2}{r}{events} & & & & & & & & & \textbf 9 \\
\midrule
 SM & $t \bar{t}$ & $11 \times 10^6$ & $4 \times 10^6$ & $1 \times 10^6$ & 440\,000 & 350\,000  & 45\,584 & 29\,942 & 1\,266 & 3 \\
\cmidrule{4-6}\cmidrule{11-11}
 \multicolumn{2}{r}{events} & & & & & & & & & \textbf 1 \\
\bottomrule
\end{tabular}
\caption{Cut flow of semi-leptonic analysis for different signals and relevant background at $\sqrt{s} = \unit[7]{TeV}$. The remaining signal and background events,
scaled to an integrated luminosity of $\unit[20]{fb^{-1}}$, are printed in bold.
The cut flow for Spectrum~I is shown separately for each different production channel.
}
\label{tab:cutflow3}
\end{table}
The resulting cut flow is displayed in Table~\ref{tab:cutflow3}.
\begin{figure}
\centering
\footnotesize
\psfrag{Events}{\small Events}
\psfrag{etmis}{\small\hspace{-2em}$\slashed E_T$ [\unit{GeV}]}
\psfrag{SM}{\small SM}
\psfrag{Spectrum I}{\small Spectrum I}
\psfrag{HH50p}{\small HH$50'$}
\psfrag{0}{0}
\psfrag{1}{1}
\psfrag{10}{10}
\psfrag{100}{100}
\psfrag{200}{200}
\psfrag{300}{300}
\psfrag{400}{400}
\psfrag{500}{500}
\psfrag{600}{600}
\psfrag{700}{700}
\psfrag{800}{800}
\psfrag{900}{900}
\psfrag{1000}{1\,000}
\includegraphics[width=0.6\textwidth]{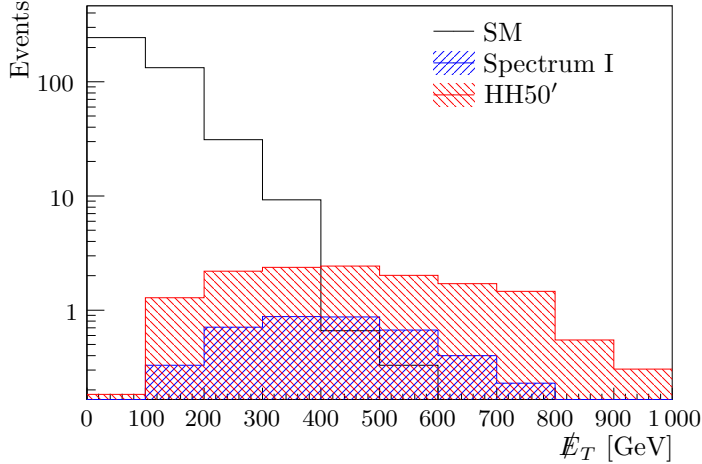}
\caption{$\slashed E_T$ distribution in the semi-leptonic analysis before the final $\slashed E_T$ cut.
SM events are black, events in Spectrum I are blue and events in HH$50'$ are red.
}
\label{fig:etmis distribution}
\end{figure}
As advertised, the number of leptonic events to survive the cuts is not significantly above the SM background, whereas a significant number of events survive in HH50 and HH$50'$ (cf.~Figure~\ref{fig:etmis distribution}). This set of cuts therefore serves to discriminate between our model and CMSSM-like models.

\subsection{Model discrimination: a simplified model}

The analysis of Section~\ref{sec:sls} relies on the presence of intermediate states
(in the case of HH50 and HH$50'$, the wino-like $\chi^\pm_1$ and $\chi^0_2$) whose decay into the LSP produces isolated leptons.
In models with non-unified gaugino masses, the LSP could still be bino-like while all remaining charginos and neutralinos are much heavier.
Can we still distinguish our model from a generic model with a comparably heavy $\tilde t_1$ and only a light bino LSP below it?
It turns out that this is rather more difficult, but still possible.

The simplified model in Table~\ref{tab:spectrum} has been designed to reproduce the relevant collider signals. We use the production cross section of stop pairs taken from Spectrum I.
The only active states are a moderately heavy $\tilde t_1$ and a light bino-like $\chi^0_1$.
Stops that are produced in pairs will decay as $\tilde t_1\into t \chi^0_1$, with the $t$ further decaying into $bW$.
The signature is therefore $b$-jets and missing energy. A similar decay chain is also open in our model (as in the lower branch in Figure~\ref{fig:stoppair}).
However, in our model about \unit[50]{\%} of the stops will decay directly into $b$ quarks and missing energy (as in the upper branch).
These latter events will produce slightly harder $b$-jets than those involving an intermediate top.

To select the stop pair production channel in our model, we impose a series of simple cuts.
At the pre-selection cut level, we select event with at least two and at most four high-energetic jets with $p_T$ larger than \unit[100]{GeV}, similar to the all-hadronic analysis:
\begin{itemize}
 \item $1<N(j)<5$ , where \quad $p_T(j) > \unit[100]{GeV}$,
 \item $\slashed{E}_T > \unit[50]{GeV}$.
\end{itemize}

Heavy squarks and gluinos will decay via long decay chains, typically giving rise to a large number of high-energetic jets.
Therefore, we select events with exactly two high-energetic jets in order to single out stop pair production.
Furthermore, we demand that at least one of these jets is a $b$-jet:
\begin{itemize}
 \item $N(j) = 2$, where \quad $p_T(j) > \unit[100]{GeV}$,
 \item $N(b\text{-jets})\geq 1$.
\end{itemize}
The invariant mass of the 2-jet system originating in such decays is sensitive to the masses of the parent particles. We select events with relatively small 2-jet transverse mass:
\begin{itemize}
\item $m^T_{jj} \equiv \sqrt{ 2 p_T(j_1) p_T(j_2) \left (1 - \cos \Delta \phi( j_1, j_2) \right) } < \unit[500]{GeV}$
\end{itemize}
In order to suppress the Standard Model background we employ following cuts:
\begin{itemize}
\item $\PTj > \unit[400]{GeV}$,
\item $\Delta \phi \left(\slashed E_T\, ,j_2  \right) > 1 $,
\item $\slashed{E}_T > \unit[400]{GeV}$,
\item $N(l) = 0$.
\end{itemize}
Missing transverse energy in QCD di- and multi-jet events can only appear due to the mismeasurement of one of the jets. We assume that, in events with very large missing transverse energy and exactly two high-energetic jets, the mismeasured jet is the next-to-leading one. We therefore expect that no QCD event will survive the cuts on $\Delta \phi \left(\slashed E_T \, ,j_2  \right)$ and $\slashed E_T$.
The resulting cut flow is displayed in Table~\ref{tab:cutflow4}.

\begin{table}
\centering
\begin{tabular}{ l l @{} r r r @{ } r @{ } r @{ } r @{ } r @{ } r @{ } r @{ } r }
\toprule
 & & \multicolumn{1}{l}{\hspace*{.8em}before} & \multicolumn{2}{c}{pre-cuts} & & & & & & & \\
\cmidrule{4-5}
 & & \multicolumn{1}{l}{\hspace*{.8em}cuts} & \multicolumn{1}{c}{$N(j)$} & \multicolumn{1}{c}{$\slashed E_T$} & \multicolumn{1}{c}{$N(j)$} & \multicolumn{1}{c}{$b$-tag} & \multicolumn{1}{c}{$m^T_{jj}$} & \multicolumn{1}{c}{$\PTj$} & \multicolumn{1}{c}{$\Delta \phi$} & \multicolumn{1}{c}{$\slashed E_T$} & \multicolumn{1}{c}{$N(l)$} \\
\midrule
 \multirow{6}{*}{\begin{sideways}Spectrum~I\end{sideways}}& $\tilde{q} \tilde{q}^*$ & 720 & 569 & 555 & 71 & 12 & 3 & 3 &2 & 2 & 2 \\
 & $\tilde{q} \tilde{q}$ & 7\,660 & 6\,416 & 6\,329 & 978 & 179 & 55 & 53 &48 & 33 & 24 \\
 & $\tilde{t} \tilde{t}^*$ & 11\,220 & 8\,909 & 8\,729 & 6\,093 & 3\,158 & 1\,928 & 1\,378 & 1\,238 & 637 & 575 \\
 & $\tilde{b} \tilde{b}^*$ & 1\,200 & 993 & 983 & 651 & 332 & 152 & 125 &116 & 72 & 63 \\
 & $\tilde{g} \tilde{g}$ & 5\,800 & 6\,478 & 4\,622 & 658 & 348 & 144 & 115 &104 & 78 & 58 \\
 & $\tilde{g} \tilde{q}$ & 16\,900 & 13\,425 & 13\,257 & 1\,803 & 684 & 243 & 201 &178 & 121 & 77 \\
\cmidrule{4-5}\cmidrule{12-12}
 \multicolumn{2}{r}{events} & & & & & & & & & & \textbf 8 \\
\midrule
 \multicolumn{2}{l}{simplified} & 11\,220 & 8\,179 & 7\,986 & 5\,328 & 2\,107 & 1\,339 & 782 & 666 & 316 & 243 \\
\cmidrule{4-5}\cmidrule{12-12}
 \multicolumn{2}{r}{events} & & & & & & & & & & \textbf 2 \\
\midrule
 \multirow{2}{*}{\begin{sideways}SM\end{sideways}} & $t \bar{t}$ & $1 \times 10^7$ & $3 \times 10^6$ & $1 \times 10^6$ & 739\,752 & 290\,416 & 268\,254 & 34\,062 & 8\,669 & 34 & 16 \\
 & $t$ & $1.7 \times 10^6$ & 160\,197 & 23\,773 & 21\,234 & 6\,858 & 6\,330 & 907 & 176 & 6 & 3 \\
\cmidrule{4-5}\cmidrule{12-12}
 \multicolumn{2}{r}{events} & & & & & & & & & & \textbf 8 \\
\bottomrule
\end{tabular}
\caption{Cut flow of the analysis in which we examine the possibility to distinguish $\tilde t$ decays via bino-like neutralinos from decays via higgsino-like neutralinos at $\sqrt{s} = \unit[7]{TeV}$.
The remaining signal and background events,
scaled to an integrated luminosity of $\unit[20]{fb^{-1}}$, are printed in bold.
}
\label{tab:cutflow4}
\end{table}

Evidently, these cuts can discriminate between Spectrum~I and the simplified model.
Of course the latter is not a realistic scenario, and in a fully-fledged model cascade decays of heavier states may also be relevant.
However, since the cuts single out the stop pair production channel in our model quite efficiently, it seems reasonable to expect that this remains true for a generic full model which the simplified model is taken to represent here.
The cuts are even tight enough to remove almost all of the stop decay events in the simplified model, while leaving a substantial excess above the Standard Model background in our model (presumably coming from direct $\tilde t_1\into b\chi_1^\pm$ decays).
Note, however, that this analysis will be rather challenging with real data: Only few events survive, and the discrimination is not mainly due to a single cut, but rather to the combined effects of all of them.

\section{Conclusions}

Supersymmetry could still be just around the corner, even though the corner has now moved a bit.
Should candidate SUSY signatures be observed within the next year, it will be interesting to see how much can already be found out
about the underlying model at the early LHC.
It is therefore important to study the collider characteristics of physically well motivated scenarios. This is a complementary approach to just studying simplified models, designed for their simple collider
phenomenology, or simple ad-hoc parametrizations such as the CMSSM.

In this paper we have investigated the early-LHC prospects for the MSSM with light higgsinos, as obtained
from simple mixed gauge-gravity mediated models which are motivated by a certain class of string compactifications. For a first tentative study of their collider phenomenology, we have analyzed a particular benchmark point in some detail. With early LHC data, evidence for our model could be found in jets plus missing transverse energy searches. Moreover, with dedicated cuts and using also the leptonic search channels, it will be possible to distinguish our model from more commonly studied standard SUSY frameworks, such as the CMSSM or a bino-LSP simplified model.

Our analysis is rather crude compared to what could be done with a full detector simulation and using state-of-the-art multivariate analysis methods.
We have also restricted ourselves to a best-case scenario with a favourable choice of parameters.
The present paper should therefore be regarded as a first step to the exploration of our class of models.
It would be very interesting to refine and extend this study to the full parameter space, as far as it can be explored by early LHC, at least once there are hints for supersymmetry.
Furthermore, it should be worthwile to also assess the discovery potential at \unit[14]{TeV}.

\subsection*{Acknowledgements}

The authors thank A.~Bharucha, S.~Brensing, T.~Figy, C.~Sander, and P.~Schleper for useful discussions.

\bibliographystyle{h-physrev}
\bibliography{bibliography}

\end{fmffile}

\end{document}